\documentclass[letterpaper, 10pt, conference]{ieeeconf}

\IEEEoverridecommandlockouts
\usepackage{amsmath} 
\usepackage{amssymb} 
\usepackage{mathtools}
\usepackage[short,c2,nocomma]{optidef}
\usepackage{cite}

\usepackage{microtype}

\usepackage[linesnumbered,ruled,noend]{algorithm2e}
	\SetKw{init}{Initialization}
	\SetKw{Input}{Input}
	\SetKw{Output}{Output}

\usepackage{siunitx}
\title{\LARGE \bf
	Convergent NMPC-based Reinforcement Learning Using Deep Expected Sarsa and Nonlinear Temporal Difference Learning
}

\author{Amine Salaje, Thomas Chevet, and Nicolas Langlois%
\thanks{*This work was supported by ANR and R{\'e}gion Normandie through the HAISCoDe (ANR-20-THIA-0021) and PAMAP projects.
	For this reason and the purpose of Open Access, the authors have applied a CC BY public copyright licence to any Author Accepted Manuscript (AAM) version arising from this submission.}%
\thanks{The authors are with Universit\'e de Rouen, ESIGELEC, IRSEEM, 76000 Rouen, France
	{\tt\small amine.salaje@groupe-esigelec.org}, {\tt\small \{thomas.chevet, nicolas.langlois\}@esigelec.fr}}%
}

\begin{document}

\maketitle
\thispagestyle{empty}
\pagestyle{empty}

\begin{abstract}
	In this paper, we present a learning-based nonlinear model predictive controller (NMPC) using an original reinforcement learning (RL) method to learn the optimal weights of the NMPC scheme, for which two methods are proposed.
	Firstly, the controller is used as the current action-value function of a deep Expected Sarsa where the subsequent action-value function, usually obtained with a secondary NMPC, is approximated with a neural network (NN).
	With respect to existing methods, we add to the NN's input the current value of the NMPC's learned parameters so that the network is able to approximate the action-value function and stabilize the learning performance.
	Additionally, with the use of the NN, the real-time computational burden is approximately halved without affecting the closed-loop performance. 
	Secondly, we combine gradient temporal difference methods with a parametrized NMPC as a function approximator of the Expected Sarsa RL method to overcome the potential parameters' divergence and instability issues when nonlinearities are present in the function approximation.
	The simulation results show that the proposed approach converges to a locally optimal solution without instability problems.
\end{abstract}

\section{Introduction}

Nonlinear model predictive control (NMPC) has proven effective in addressing complex control challenges \cite{grune2017nonlinear}, particularly in systems with nonlinear dynamics and constraints, such as mobile robotics \cite{nascimento2018nonholonomic}.
It generates control signals by solving a constrained optimization problem -- the optimal control problem (OCP) -- at each time step. 
However, fine-tuning the parameters of the OCP remains a challenge, with ongoing research seeking reliable tuning methods \cite{zhang2022advances}.
Reinforcement learning (RL) offers a promising solution as it enables an agent to optimize these parameters by learning optimal value functions and policies within the scope of a Markov decision process (MDP) \cite{sutton2018reinforcement}.

In problems with large MDPs or high-dimensional state spaces, modern RL methods often use deep neural networks (DNN) to approximate value functions or policies, allowing them to handle the complexity of nonlinear systems.
However, DNN-based RL approaches often face difficulties in closed-loop stability analysis and can be challenging for formal verification and constraint management \cite{gros2019data}, thus raising concerns about ensuring safety in control systems. 
To address these issues, NMPC-based RL has been proposed and justified in \cite{gros2019data} as a promising solution.
By using the OCP as a function approximator for the optimal policy in RL, this approach naturally satisfies state and input constraints while ensuring safety requirements.

In \cite{gros2019data}, the NMPC parameters are tuned to enhance closed-loop performance using Q-learning and policy gradient RL methods. 
Since then, various studies have deployed Q-learning with MPC as a function approximator to realize this task, with notable contributions found in \cite{zanon2019practical,gros2020safe,martinsen2020combining,kordabad2021reinforcement,esfahani2021approximate}.
However, one of the current challenges of these approaches is the significant computational demand for real-time implementation \cite{kordabad2023reinforcement}.
Indeed, in the context of temporal difference (TD) learning methods, it is required to solve at least two constrained nonlinear optimization problems at each sampling instant -- one for estimating the current action-value function and another for approximating the subsequent action-value function.

To deal with this issue, a new data-driven approach was proposed in \cite{moradimaryamnegari2023data}, where a parametrized NMPC has been used as an approximator for a deep double Expected Sarsa (ES) RL method.
To estimate the current action-value function, a parametrized OCP is solved at each time step.
However, the subsequent action-value function is approximated using a target neural network (NN) instead of solving a second OCP at the next time step.
The NN is trained with inputs and outputs of the primary NMPC obtained at previous sampling times.
This approach shows a clear advantage in reducing online computation time compared to \cite{gros2019data}.
However, as the second NMPC of \cite{gros2019data} is replaced by a NN in \cite{moradimaryamnegari2023data}, the simulation results show instability and potential divergence.
As a first contribution, we propose in the present paper to tackle this problem by adding the parametrization vector of the NMPC at previous time steps to the training input of the NN, leading to a clear improvement in the convergence stability and control performance compared with \cite{moradimaryamnegari2023data}.

While simple to use, conventional TD learning methods such as Q-learning lack formal guarantees  for the closed-loop optimality of the learned policy \cite{sutton2018reinforcement}.
Furthermore, they lack convergence guarantees when using nonlinear function approximation, which may cause potential divergence of the approximator parameters.
To overcome these challenges, the authors of \cite{maei2010toward} proposed gradient temporal difference (GTD) for off-policy Q-learning methods as an effective solution, a method ensuring almost-sure convergence for any finite MDP and any smooth value function approximator to a locally optimal solution or equilibrium point. 
As a second contribution, we combine in this paper GTD methods with parametrized NMPC as a function approximator in the Expected Sarsa (ES) RL method to overcome the potential parameters divergence and instability issues.
To the best of the authors' knowledge, this is the first work where an action-value method is used to tune the NMPC approximator using GTD methods. 
The simulation results show that the proposed approach converges to a locally optimal solution without instability problems compared to previous methods.

The remainder of this article is organized as follows.
Section \ref{section:prerequisites} exposes prerequisites on predictive control and reinforcement learning.
Then, Section \ref{section:learning_nmpc} presents our contributions. 
Finally, Section \ref{section:results} showcases the efficiency of our methods on numerical simulation examples.

\section{Prerequisites on learning and control}
	\label{section:prerequisites}

We start by giving some context on the considered control strategy and background on the RL methods we use.

	\subsection{Markov decision process for RL}
		\label{subsection:MDP}

Consider the tuple $(\mathcal{S}, \mathcal{A}, \mathbb{P}, \ell, \rho, \gamma)$ which defines an MDP with continuous state space $\mathcal{S}\subseteq\mathbb{R}^n$ and action space $\mathcal{A}\subseteq\mathbb{R}^m$.
The state transition dynamics have the underlying probability law $\mathbb{P}$ such that $\mathbf{s}_{k + 1} \sim \mathbb{P}(\cdot|\mathbf{a}_k, \mathbf{s}_k)$, with $\mathbf{s}_{k + 1}$, $\mathbf{s}_k\in\mathcal{S}$ and $\mathbf{a}_k\in\mathcal{A}$.
In a control approach, these state transition dynamics can be expressed as $\mathbf{s}_{k + 1} = f(\mathbf{a}_k, \mathbf{s}_k)$ where $f$ is the discretized real system dynamics.
The remaining elements of the tuple are the reward $\ell$, or stage cost in optimal control, which is associated to each MDP transition, the initial state distribution $\rho$, and the discount factor $\gamma\in(0, 1)$ determining the importance of future rewards.

The goal of RL is to find a policy, denoted by $\boldsymbol{\pi}\colon\mathcal{S}\rightarrow\mathcal{A}$, minimizing a closed-loop performance objective \cite{sutton2018reinforcement}.
Consider then a deterministic policy delivering the control input $\mathbf{a}_k = \boldsymbol{\pi}(\mathbf{s}_k)$ resulting in state distribution $\rho^\pi$.
Given an objective $J(\boldsymbol{\pi}) = \mathbb{E}_{\mathbf{s} \sim \rho^\pi}\left[\sum_{k=0}^{K} \gamma^k \ell\left(\mathbf{s}_k, \mathbf{a}_k\right)\middle|\mathbf{a}_k=\boldsymbol{\pi}(\mathbf{s}_k)\right]$ on the expected discounted cumulative cost, with $K\in\mathbb{N}\cup\{+\infty\}$ a (possibly infinite) horizon, the RL problem aims to find the optimal policy $\boldsymbol{\pi}^*$ satisfying $\boldsymbol{\pi}^{\star} = \arg\min J(\boldsymbol{\pi})$.

The optimal action-value function $Q^{\star}(\mathbf{s}_k, \mathbf{a}_k)$ and value function $V^{\star}(\mathbf{s}_k)$ associated to the optimal policy $\boldsymbol{\pi}^*$ can be determined using the Bellman equations
\begin{equation}
	\begin{aligned}
		Q^{\star}(\mathbf{s}_k, \mathbf{a}_k) & =\ell(\mathbf{s}_k, \mathbf{a}_k) + \gamma\mathbb{E}\left[V^{\star}\left(\mathbf{s}_{k + 1}\right) \middle| \mathbf{s}_k, \mathbf{a}_k\right]\text{,}\\
		V^{\star}(\mathbf{s}_k) & = Q^{\star}\left(\mathbf{s}_k, \boldsymbol{\pi}^{\star}(\mathbf{s}_k)\right) = \min Q^{\star}(\mathbf{s}_k, \mathbf{a}_k).
	\end{aligned}
\end{equation}

	\subsection{NMPC as a function approximator}
		\label{subsection:nmpc}

As in \cite{gros2019data}, we use a parametric optimization problem as a function approximator for the reinforcement learning.
We therefore consider the parametrized NMPC 
\begin{mini!}
	{%
		\substack{\mathbf{u}_i, \mathbf{x}_i, \sigma_i\text{, }\\\forall i\in\overline{0, N}}
	}{%
		\sum_{i = 0}^{N - 1} \gamma^i \left(L\left(\mathbf{x}_i, \mathbf{u}_i, \boldsymbol{\vartheta}\right) + \boldsymbol{\omega}^{\top}\boldsymbol{\sigma}_i \right)
			\nonumber
	}{\label{eq:ocp}}{}
	\breakObjective{\hspace{4em}+ \gamma^N  \left( V (\mathbf{x}_N, \boldsymbol{\vartheta}_f) + \boldsymbol{\omega}_f^{\top}\boldsymbol{\sigma}_N \right)\label{eq:ocp_cost_function}}
	\addConstraint{%
		\mathbf{x}_0%
	}{%
		= \mathbf{s}_k\text{,}%
	}{%
		\label{eq:ocp_constraint_initial}%
	}
	\addConstraint{%
	\mathbf{x}_{i + 1}%
	}{%
		= f(\mathbf{u}_i, \mathbf{x}_i)\text{,}%
	}{%
		\hspace{-8em}\forall i\in\overline{1, N}\text{,}%
			\label{eq:ocp_constraint_dynamics}%
	}
	\addConstraint{%
		\mathbf{x}_i%
	}{%
		\in\mathcal{S}\text{,}%
	}{%
		\hspace{-8em}\forall i\in\overline{1, N}\text{,}%
			\label{eq:ocp_constraint_state}
	}
	\addConstraint{%
		\mathbf{u}_i%
	}{%
		\in\mathcal{A}\text{,}%
	}{%
		\hspace{-8em}\forall i\in\overline{0, N - 1}\text{,}%
			\label{eq:ocp_constraint_control}
	}
	\addConstraint{%
		g(\mathbf{x}_i) + \theta_c %
	}{%
		\leq \boldsymbol{\sigma}_i\text{,}%
	}{%
		\hspace{-8em}\boldsymbol{\sigma}_i\geq 0\text{, }\forall i\in\overline{0, N}\text{.}%
			\label{eq:ocp_constraint_obstacle}
	}
\end{mini!}
In this formulation, $\forall i\in\overline{0,N}$, the variables $\mathbf{u}_i$, $\mathbf{x}_i$, and $\boldsymbol{\sigma}_i$ represent the primal decision variables, $L$ the NMPC stage cost and $V$ the terminal cost. 
The prediction horizon $N\in\mathbb{N}$ may be shorter than $K$ from the performance measure $J(\boldsymbol{\pi})$, and $\gamma$ is the discount factor referenced in Paragraph \ref{subsection:MDP}.
The function $f$ denotes the deterministic dynamic model driven by the NMPC scheme \eqref{eq:ocp} as in Paragraph \ref{subsection:MDP}, while the function $g$ imposes additional constraints on the state at each time step over the prediction horizon $N$.
The variables $\boldsymbol{\sigma}_i$, for $i\in\overline{0,N}$, are slack variables, introduced to relax state constraints, with their violation penalized by the weights $\boldsymbol{\omega}$ and $\boldsymbol{\omega}_f$, preventing the potential infeasibility of the NMPC due to the additional constraints.
Finally, $\boldsymbol{\vartheta}$ and $\boldsymbol{\vartheta}_f$ are vectors of NMPC weights, usually selected \textit{a priori} based, e.g., on the dynamics $f$ or the control objective for the system.
In this paper, these weights are treated as parameter vectors to be adjusted through an RL algorithm.

Indeed, an RL algorithm can advantageously adjust the parameters of the NMPC cost function, model, and constraints in a way to enhance closed-loop performance.
In theory, as stated in \cite[Theorem 1]{gros2019data}, under specific assumptions and with a sufficiently expressive parametrization, the MPC scheme can approximate the optimal policy $\boldsymbol{\pi}^{\star}$.

When solving \eqref{eq:ocp} at time $k\in\mathbb{N}$, we obtain the optimal control sequence $\mathbf{u}^{\star}(\mathbf{s}_k, \boldsymbol{\theta}) = \{\mathbf{u}^{\star}_0(\mathbf{s}_k, \boldsymbol{\theta}),\ldots,\mathbf{u}^{\star}_{N-1}(\mathbf{s}_k, \boldsymbol{\theta})\}$, with $\boldsymbol{\theta} = \begin{bmatrix}\boldsymbol{\vartheta} & \boldsymbol{\vartheta}_f\end{bmatrix}$ driving the system with dynamics \eqref{eq:ocp_constraint_dynamics} towards its objective over the prediction horizon $N$.

	\subsection{Expected Sarsa with NMPC as function approximator}
		\label{subsection:double_expected_sarsa}

ES is a TD control method used in model-free reinforcement learning \cite{sutton2018reinforcement}.
To handle large MDPs, the ES algorithm employs function approximators parametrized by unknown parameters $\boldsymbol{\theta}$ to approximate the optimal action-value function $Q^{\star}$.

Under some assumptions given in \cite{gros2019data}, the NMPC scheme \eqref{eq:ocp} approximates the optimal action-value function $Q^{\star}$.
Thus, for any $k\in\mathbb{N}$, we define
	\begin{equation}\label{eq:mpc_qfunction}
		Q_{\boldsymbol{\theta}}\left(\mathbf{s}_k, \mathbf{a}_k\right) = \eqref{eq:ocp}\text{,}
	\end{equation}
an approximator of the optimal action-value function $Q^{\star}$.
Unlike Q-learning,  when combining NMPC with the ES method for on-policy learning, the action-value function \eqref{eq:mpc_qfunction} does not need to be evaluated at the first action $\mathbf{u}_0= \mathbf{a}_k$ \cite{moradimaryamnegari2022model} (the behavioral policy is the same as the target policy).
Furthermore, we keep the same notation for $Q_{\boldsymbol{\theta}}\left(\mathbf{s}_k, \mathbf{a}_k\right) $ (rather than $ V(\mathbf{s}_k)$) with respect to our comparison with \cite{moradimaryamnegari2022model,moradimaryamnegari2023data}.

The policy $\boldsymbol{\pi_{\theta}}(\mathbf{s}_k)$ and the action $\mathbf{a}_k$ are obtained by solving \eqref{eq:ocp}, so that, for any $\mathbf{s}_k\in\mathcal{S}$, 
\begin{align} 
	\label{eq:policy_function}
		\boldsymbol{\pi}_{\boldsymbol{\theta}}(\mathbf{s}_k) & = \arg\min_{\mathbf{a}_k\in\mathcal{A}} Q_{\boldsymbol{\theta}}(\mathbf{s}_k, \mathbf{a}_k) = \mathbf{u}_0^{\star}(\mathbf{s}_k, \boldsymbol{\theta})
	\text{,}\\
	\label{eq:action}
		\mathbf{a}_k  & = \boldsymbol{\pi_{\theta}}(\mathbf{s}_k) + c_\varepsilon^{k}\boldsymbol{\varepsilon}
\end{align}
satisfying the Bellman equation \cite{sutton2018reinforcement}. 
Here, $\boldsymbol{\varepsilon} \sim\mathcal{N}(0, \num{1})$ represents a Gaussian term introducing weighted exploration of the continuous action space $\mathcal{A}$.
This exploration decreases exponentially during training due to the factor $c_{\varepsilon}\in(0, 1)$.
During the learning process, the action $\mathbf{a}_k$ defined by \eqref{eq:action}, is applied to the system at time step $k$, resulting in $\mathbf{s}_{k + 1}$.
 
To simplify the notations, the time dependence is dropped for states $\mathbf{s}\in\mathcal{S}$ and actions $\mathbf{a}\in\mathcal{A}$.
Then, $\mathbf{s}^+\in\mathcal{S}$ denotes the state at the next time instant.

In order to find $Q^{\star}$, we characterize the optimal parameters $\boldsymbol{\theta}^{\star}$ as those minimizing the least-squares problem
\begin{equation}\label{eq:Q_objective}
	\boldsymbol{\theta}^{\star} = \arg \min_{\boldsymbol{\theta}} \mathbb{E}_{\boldsymbol{\pi}_{\boldsymbol{\theta}}}\left[\frac{1}{2}\left(Q^{\star}(\mathbf{s}, \mathbf{a})-Q_{\boldsymbol{\theta}}(\mathbf{s}, \mathbf{a})\right)^2\right].
\end{equation}
Since the optimal $Q^{\star}$ is usually unknown, we cannot solve \eqref{eq:Q_objective} directly.
To do so, we use here the continuous ES algorithm \cite{moradimaryamnegari2022model,van2009theoretical}.

The parametrized action-value function $Q_{\boldsymbol{\theta}}$ is estimated using the NMPC scheme \eqref{eq:ocp} parametrized by the vector $\boldsymbol{\theta}$.
At each sampling time, the parameters are updated based on the TD error
\begin{equation}\label{eq:sarsa_tderror}
	\delta = \ell\left(\mathbf{s}, \mathbf{a}\right)  + \gamma Q_{\boldsymbol{\theta}}\left(\mathbf{s}^+, \mathbf{a}^+\right) - Q_{\boldsymbol{\theta}}\left(\mathbf{s}, \mathbf{a}\right)
\end{equation}
where $\mathbf{a}^+$ is obtained by applying $\mathbf{a}$, obtained as in \eqref{eq:action} by solving the OCP \eqref{eq:ocp} for state $\mathbf{s}$ (behavioral policy), to the controlled system and solving again the OCP \eqref{eq:ocp} for the resulting state $\mathbf{s}^+$ (target policy).
The parameter updates are determined by minimizing the mean-square TD error 
\begin{equation}
	\min_{\boldsymbol{\theta}} \mathbb{E}_{\boldsymbol{\pi}_{\boldsymbol{\theta}}}\left[\frac{1}{2}\left(\ell\left(\mathbf{s}, \mathbf{a}\right)  + \gamma Q_{\boldsymbol{\theta}}\left(\mathbf{s}^+, \mathbf{a}^+\right)- Q_{\boldsymbol{\theta}}(\mathbf{s}, \mathbf{a})\right)^2\right].
\end{equation}
We apply the semi gradient step to perform this minimization, thus giving the parameters' update rule
\begin{equation} \label{eq:expected_sarsa_update_parameters}
	\boldsymbol{\theta}\leftarrow \boldsymbol{\theta} + \alpha\delta\nabla_{\boldsymbol{\theta}}Q_{\boldsymbol{\theta}}(\mathbf{s}, \mathbf{a})
\end{equation}
where $ \alpha > 0 $ is the learning rate. By updating the unknown parameters, the action-value functions can converge towards its optimal value.
The gradient $\nabla_{\boldsymbol{\theta}}Q_{\boldsymbol{\theta}}(\mathbf{s}, \mathbf{a})$ is obtained from sensitivity analysis.
For more details the reader is referred to \cite{gros2019data}.

\section{Proposed methods}
	\label{section:learning_nmpc}

This section presents the main contributions of this paper, aiming to improve the overall performances of the learning algorithm.
Our methods are then summarized in Algorithm \ref{alg:des_parameter_tuning}.

	\subsection{Reducing computational complexity of NMPC-based ES}
		\label{subsection:regularized_deep_expected_sarsa}

		\subsubsection{\textbf{Problem definition}}

In \cite{gros2019data} and \cite{moradimaryamnegari2022model}, the parameter vector $\boldsymbol{\theta}$ is tuned using Q-learning and an ES algorithm, respectively.
In both cases, it is required to solve, at each learning step, a first OCP of the form \eqref{eq:ocp} to obtain $Q_{\boldsymbol{\theta}}(\mathbf{s}, \mathbf{a})$ as well as a second OCP to obtain the subsequent action-value function $Q_{\boldsymbol{\theta}}\left(\mathbf{s}^+, \mathbf{a}^+\right)$.
These two action-value functions are then used to compute the TD error to be minimized.
However, solving two constrained nonlinear optimization problems at each time step significantly increases the real-time computational complexity, resulting in longer training times. 
This issue worsens with large prediction horizons $N$, when the OCP \eqref{eq:ocp} involves multiple constraints, etc. 
To address this problem, \cite{moradimaryamnegari2023data} replace the second OCP of the aforementioned methods by a neural network approximator.
However, this method may diverge due to an inaccurate approximation of the subsequent action-value function.

		\subsubsection{\textbf{NMPC-based off-policy regularized deep ES (RDES)}}

To better approximate the subsequent action-value function obtained with a second NMPC \cite{moradimaryamnegari2022model} using an NN, we propose to adjust the input data used to train the NN.
The idea is to learn the action-value function $Q_{\boldsymbol{\theta}}$ of \eqref{eq:mpc_qfunction} using a neural network $Q_{\boldsymbol{\varpi}_{\text{NN}}}$ parametrized by the weights $\boldsymbol{\varpi}_{\text{NN}}$ so that $Q_{\boldsymbol{\varpi}_{\text{NN}}}(\mathbf{s},\mathbf{a},\boldsymbol{\theta}) \approx Q_{\boldsymbol{\theta}}(\mathbf{s},\mathbf{a})$ .
To do so, the NN training input should be chosen carefully.

In order to train the NN, we store, at each learning step, the tuple $\{\mathbf{s}, \mathbf{a}, \boldsymbol{\theta},  Q_{\boldsymbol{\theta}}\left(\mathbf{s}, \mathbf{a}\right)\}$ into a replay buffer $\mathcal{D}$.
The current state $\mathbf{s}$, the action $\mathbf{a}$ and the parametrization vector $\boldsymbol{\theta}$ are the NN's inputs while $Q_{\boldsymbol{\theta}}\left(\mathbf{s}, \mathbf{a}\right)$ is the regression target.
Then, a mini-batch $\mathcal{B}\subset\mathcal{D}$ of $n\in\mathbb{N}$ elements is randomly sampled and we update $\boldsymbol{\varpi}_{\text{NN}}$ using the Adam algorithm \cite{kingma2014adam} with a learning rate $\zeta$ to minimize the cost function
	\begin{equation} \label{eq:cost_training_nn}
		C_{\text{NN}}(\theta) = \frac{1}{n}\sum_{\{\boldsymbol{\varsigma},\boldsymbol{\eta},\boldsymbol{\vartheta},q\}\in\mathcal{B}}\left(Q_{\boldsymbol{\varpi}_{\text{NN}}}(\boldsymbol{\varsigma}, \boldsymbol{\eta}, \boldsymbol{\vartheta}) - q\right)^2
	\end{equation}
where $Q_{\boldsymbol{\varpi}_{\text{NN}}}(\boldsymbol{\varsigma}, \boldsymbol{\eta}, \boldsymbol{\vartheta})$ for $\{\boldsymbol{\varsigma}, \boldsymbol{\eta}, \boldsymbol{\vartheta}, q\}\in\mathcal{B}\subset\mathcal{D}$ is the output of the NN with parameters $\boldsymbol{\varpi}_{\text{NN}}$ and inputs $\boldsymbol{\varsigma}$, $\boldsymbol{\eta}$, and $\boldsymbol{\vartheta}$, whereas $q$ is the regression target.

After the NN's update, the subsequent action-value function in \eqref{eq:sarsa_tderror} is predicted by passing to the NN the next state $\mathbf{s}^+$ and the new target policy defined as $\mathbf{a}^+ = \boldsymbol{\pi}_{\boldsymbol{\theta}}(\mathbf{s}^+) = \mathbf{u}_1^{\star}(\mathbf{s}, \boldsymbol{\theta})$ the optimal value of $\mathbf{u}_1$, the second input vector over the prediction horizon when solving the primary NMPC \eqref{eq:policy_function}.
By doing so, only one constrained optimization problem needs to be solved at each time step as well as the NN optimization process.
Finally, we define the new TD error to be minimized, at each learning step, by
	\begin{equation} \label{eq:ddes_modified_td_error}
		\varrho = \ell\left(\mathbf{s}, \mathbf{a}\right)  + \gamma Q_{\boldsymbol{\varpi}_{\text{NN}}}(\mathbf{s}^+,\mathbf{a}^+, \boldsymbol{\theta}) - Q_{\boldsymbol{\theta}}\left(\mathbf{s}, \mathbf{a}\right)\text{.}
	\end{equation}
We note that even when replacing the subsequent action-value function by the NN approximator, the constraint satisfaction is always maintained since the  policy is derived from the OCP scheme \eqref{eq:ocp}.
Compared with \cite{moradimaryamnegari2022model}, our contribution then consists in adjusting the input data used to train the NN.
We incorporate the current value of the parametrization vector $\boldsymbol{\theta}$ of the action-value function defined in \eqref{eq:mpc_qfunction} to the NN's input.
This is based on the assumption that the approximated subsequent action-value function should depend on $\boldsymbol{\theta}$, as the current action-value function takes vector $\boldsymbol{\theta}$ as input at each time step.
Thus, by adding the parameter vector $\boldsymbol{\theta}$ to the training data, the NN gains more knowledge about the behavior of the primary NMPC \eqref{eq:ocp}.
Therefore, stabilizing the learning performance and improving the approximation of the action-value function, which is usually computed by solving a second NMPC, under the assumption
	\begin{equation} \label{eq:nn_convergence}
		Q_{\boldsymbol{\varpi}_{\text{NN}}}(\mathbf{s},\mathbf{a}, \boldsymbol{\theta})  \rightarrow Q_{\boldsymbol{\theta}}\left(\mathbf{s}, \mathbf{a}\right)\text{.}
	\end{equation}

	\subsection{Nonlinear temporal difference learning}
			\label{subsection:GTD}

		\subsubsection{\textbf{Problem definition}}

Traditional TD methods like Q-learning and ES have been successfully applied with function approximation in various fields.
However, it is well-known that these algorithms can become unstable, particularly when nonlinear function approximation is used, which can lead to parameter divergence in the approximator.
To address this issue, we build on the results from \cite{maei2009convergent, maei2010toward}, and our idea is to use a GTD approach to address instability and convergence issues when using NMPC as function approximator of RL methods.
Indeed, the GTD methods are proven to converge almost surely for any MDP, even with nonlinear function approximators, to a locally optimal solution or equilibrium point.

		\subsubsection{\textbf{Convergent NMPC-based gradient ES (GES)}}

In TD learning control methods, most works use an objective function representing how closely the approximate action-value function satisfies the Bellman equations.
The true action-value function satisfies the Bellman equation in model-free RL
	\begin{equation}\label{eq:belman_op}
		\Gamma^\pi Q(\mathbf{s},\mathbf{a}) = \mathbb{E}_{\pi}\left[ \ell(\mathbf{s},\mathbf{a}) +\gamma  Q(\mathbf{s}^+,  \mathbf{a}^+)\right]
	\end{equation}
where $\Gamma^\pi$ is known as the Bellman operator. 
In GTD approaches, the objective function that measure how closely the approximation $Q_{\boldsymbol{\theta}}(\mathbf{s}, \mathbf{a})$ satisfies the Bellman equation \eqref{eq:belman_op} is known as the mean-square projected Bellman error (MSPBE), where the goal is to minimize
	\begin{equation}
		\gimel(\boldsymbol{\theta}) = \left\|\Pi(\Gamma^{\pi_{\boldsymbol{\theta}}} Q_{\boldsymbol{\theta}}) - Q_{\boldsymbol{\theta}}\right\|_{\rho^{\pi}}^2
	\end{equation}
using approximate gradient decent, $\Pi$ is a projection operator which projects action-value functions into the space $\mathcal{F} = \left\{Q_{\boldsymbol{\theta}}:\boldsymbol{\theta} \in \mathbb{R}^d\right\}$ w.r.t $\|\cdot\|_{\rho^{\pi}}$ : $\Pi(\hat{Q}) = \arg\min_{f \in \mathcal{F}}\|\hat{Q}-f\|_{\rho^{\pi}}$.

Following \cite{maei2010toward}, the projected objective can be written in terms of expectations as 
\begin{equation}
	\gimel (\boldsymbol{\theta}) = \mathbb{E}[\delta\boldsymbol{\phi}]^{\top} \mathbb{E}\left[\boldsymbol{\phi}\boldsymbol{\phi}^{\top}\right]^{-1} \mathbb{E}[\delta\boldsymbol{\phi}],
\end{equation}
where $\delta$ is the TD error as in \eqref{eq:sarsa_tderror}, and $\boldsymbol{\phi}$ is the feature vector used in approximating the action-value function. 
In \cite{maei2010toward}, the authors used  an unrestricted linear form for the action-value function approximation, specifically  $Q_{\boldsymbol{\theta}}(\mathbf{s}, \mathbf{a}) = \boldsymbol{\phi}^{\top} \boldsymbol{\theta}$.
However, in this paper, we use a nonlinear function approximation defined by the OCP scheme \eqref{eq:ocp}. 
Therefore, we represent $\boldsymbol{\phi}$ as the gradient of the action-value function, expressed as $\boldsymbol{\phi} = \nabla_{\boldsymbol{\theta}}Q_{\boldsymbol{\theta}}(\mathbf{s}, \mathbf{a})$.

Finally, by applying the negative gradient on the MSPBE objective function, and introducing a new set of weights $\mathbf{w}$ making use of the weight doubling trick \cite{sutton2009fast,maei2009convergent,maei2010toward}, we get the update rules
	\begin{equation}\label{eq:GTD_update}
		\begin{aligned}
			\boldsymbol{\theta} & \leftarrow \boldsymbol{\theta} + \alpha\left[\delta\boldsymbol{\phi} - \gamma\left(\mathbf{w}^{\top}\boldsymbol{\phi}\right) \boldsymbol{\phi}^{+}\right]\text{,}\\
			\mathbf{w} & \leftarrow \mathbf{w}+\beta\left[\delta - \boldsymbol{\phi}^{\top}\mathbf{w}\right]\boldsymbol{\phi}\text{,}
		\end{aligned}
	\end{equation}
where $\boldsymbol{\phi}^{+}$ is the subsequent action-value features $\nabla_{\boldsymbol{\theta}}Q_{\boldsymbol{\theta}}(\mathbf{s}^+,\mathbf{a}^+)$.
Under some assumptions \cite{maei2010toward} on the positive step sizes $\alpha,\beta > 0$, it follows that $\boldsymbol{\theta}$ converges almost surely to a stable equilibrium point. 

\begin{algorithm}[!t]
\DontPrintSemicolon
\Input: $\boldsymbol{\theta}$, $\mathbf{w}$, $\boldsymbol{\varpi}_{\text{NN}}$, dynamical system $f$, NMPC \eqref{eq:ocp}\;
\Output: $\boldsymbol{\theta}^{\star}$\;
\init: $\mathcal{D} \leftarrow \emptyset$\;
\For{$k \leftarrow 1$ \KwTo $N_{\text{episode}}$}{
	$\mathbf{s}\leftarrow \mathbf{s}_0$\;
	\For{$t\leftarrow 0$ \KwTo $K$}{
		Solve \eqref{eq:ocp} in $\mathbf{s}$ to obtain $\mathbf{a}$, and $Q_{\boldsymbol{\theta}}\left(\mathbf{s}, \mathbf{a}\right)$\;
		Get the action $\mathbf{a}$ from \eqref{eq:action}\;
		Observe $\mathbf{s}^{+} = f(\mathbf{a}, \mathbf{s})$ and get the reward $\ell$\; %
		\If{RDES used}{
		Store  $\mathcal{D} \leftarrow \mathcal{D}\cup\{\mathbf{s}, \mathbf{a}, \boldsymbol{\theta}, Q_{\boldsymbol{\theta}}\left(\mathbf{s}, \mathbf{a}\right)\}$\;
		\If{$\operatorname{Card}(\mathcal{D})\geq n$ \label{alg_line:mini_batch}}{
			Randomly sample mini-batch of $n$ elements of $\mathcal{D}$ and get $\boldsymbol{\varpi}_{\text{NN}}$  \eqref{eq:cost_training_nn}\;
		}
		
		Compute the TD error $\delta$ with \eqref{eq:ddes_modified_td_error}\;
		Update  parameters $\boldsymbol{\theta}$ by \eqref{eq:expected_sarsa_update_parameters}\;
		}
		\ElseIf{GES used} {
		Solve \eqref{eq:ocp} in $\mathbf{s}^{+}$ to obtain $Q_{\boldsymbol{\theta}}\left(\mathbf{s}^+, \mathbf{a}^+\right)$\;
		Compute the TD error $\delta$ with \eqref{eq:sarsa_tderror}\;
		Update  parameters $\boldsymbol{\theta}$ and $\mathbf{w}$ by \eqref{eq:GTD_update}\; 
		}
		$\mathbf{s}\leftarrow \mathbf{s}^{+} $\;
	}
}
\caption{Learning algorithm.}
\label{alg:des_parameter_tuning}
\end{algorithm}

\section{Numerical applications}
	\label{section:results}

In this section we apply both our RL-based methods to the case of setpoint tracking problem with static obstacle avoidance for a diff-drive mobile robot.
These simulations aim to show the clear improvement in the methods we propose compared with the methods from \cite{moradimaryamnegari2022model} and \cite{moradimaryamnegari2023data}.

	\subsection{Simulation model}
		\label{subsection:model}

We consider a diff-drive robot with continuous-time dynamics \cite{sani2021dynamic}
	\begin{subequations} \label{eq:true_model_robot}
		\begin{align}
			\dot{x} & = v\cos\varphi\text{,}\\
			\dot{y} & = v\sin\varphi\text{,}\\
			\dot{\varphi} & = \nu\text{.}
		\end{align}
	\end{subequations}
The robot's state is defined by its coordinates $(x, y)$ in a 2D plane and its orientation $\varphi$ with inputs its linear and angular speeds $v$ and $\nu$.
For training, we discretize the dynamics \eqref{eq:true_model_robot} with the Runge-Kutta 4 method and a sampling period $T_s = \SI{0.2}{\second}$, so that
\begin{equation} \label{eq:discrete_time_robot_model}
	\mathbf{s}_{k + 1} = f(\mathbf{u}_k, \mathbf{s}_k),
\end{equation}
where $\mathbf{s}_k^{\top} = \begin{bmatrix}x_k & y_k & \varphi_k\end{bmatrix}$ and $\mathbf{u}_k^{\top} = \begin{bmatrix}v_k & \nu_k\end{bmatrix}$.

The robot objective is to find the shortest path to reach an equilibrium point $(\mathbf{s}^{\text{ref}}, \mathbf{u}^{\text{ref}})$ of the dynamics \eqref{eq:discrete_time_robot_model}, starting from a given initial position $\mathbf{s}_0$.
Here $\mathbf{s}^{\text{ref}}$ denotes the desired target state vector. 
Along its trajectory, the robot must maintain a desired safe distance from obstacles.
To this end, the robot is controlled using an NMPC as defined in \eqref{eq:ocp}. 
The cost function \eqref{eq:ocp_cost_function} is designed such that, for any $i\in\overline{0,N - 1}$,
\begin{align*}\label{eq:NMPC_cost}
	L(\mathbf{s}_i, \mathbf{u}_i, \boldsymbol{\theta}) & = \left\|\mathbf{s}_i -\mathbf{s}^{\text{ref}}\right\|_{\operatorname{diag}(\boldsymbol{\vartheta}_x)^2}^2	+ \left\|\mathbf{u}_i - \mathbf{u}^{\text{ref}}\right\|_{\operatorname{diag}(\boldsymbol{\vartheta}_u)^2}^2\text{,}\\
	\Gamma (\mathbf{s}_N, \boldsymbol{\theta}) & = \left\|\mathbf{s}_N - \mathbf{s}^{\text{ref}}\right\|_{\operatorname{diag}(\boldsymbol{\vartheta}_f)^2}^2
\end{align*}
where $\operatorname{diag}(\boldsymbol{\theta})$ is a diagonal matrix having for elements the components of $\boldsymbol{\theta}$, with $\boldsymbol{\vartheta}_x = \begin{bmatrix}\theta_x & \theta_y & \theta_{\varphi}\end{bmatrix}$, $\boldsymbol{\vartheta}_u = \begin{bmatrix}\theta_v & \theta_{\nu}\end{bmatrix}$, and $\boldsymbol{\vartheta}_f = \begin{bmatrix}\theta_{x_f} & \theta_{y_f} & \theta_{\varphi_f}\end{bmatrix}$, and $\left\|\mathbf{x}\right\|_Q^2 = \mathbf{x}^{\top}Q\mathbf{x}$.

The action space is $\mathcal{A} = [\SI{-0.6}{\meter\per\second}, \SI{0.6}{\meter\per\second}]\times[-\pi/2,\pi/2]$.
The state space is $\mathcal{S} = [\SI{-9}{\meter}, \SI{3}{\meter}]\times[\SI{-9}{\meter}, \SI{3}{\meter}]\times\mathbb{R}$.
The robot has to stay at a given distance from 4 obstacles by satisfying the inequality constraint \eqref{eq:ocp_constraint_obstacle}, defined, for $n\in\overline{1,4}$, as
\begin{equation} \label{eq:obstacle_constraint}
	\underbrace{1 - 4\frac{(x_i - x_{\text{obs},n })^2 + (y_i - y_{\text{obs},n})^2} {(d_{\text{rob}} + d_{\text{obs}})^2}}_{\Xi(\mathbf{x}_i)} + \theta_c \leq 0,
\end{equation}
where $(x_{\text{obs},n}, y_{\text{obs},n})$ is the center of the $n$-th obstacle, while $d_{\text{rob}} = \SI{0.5}{\meter}$ and $d_{\text{obs}} = \SI{2}{\meter}$ represent the diameters of a safety circle surrounding  the robot and an obstacle, respectively.
Parameter $\theta_c$ serves as a tuning variable, adjusting the intensity of the collision avoidance constraints.
Overall, the parameters vector is $\boldsymbol{\theta} = \begin{bmatrix}\theta_x & \theta_y & \theta_{\varphi} & \theta_v & \theta_{\nu} & \theta_{x_f} & \theta_{y_f} & \theta_{\varphi_f} & \theta_c\end{bmatrix}$.

To train the RL agent, the robot's initial state is set as $\mathbf{s}_0^{\top} = \begin{bmatrix}\SI{-2.5}{\meter} & \SI{1.5}{\meter} & \SI{0}{\radian}\end{bmatrix}$, and the target state is defined as $\mathbf{s}^{\text{ref}} = \begin{bmatrix}\SI{8.5}{\meter} & \SI{2}{\meter} & \pi/2\end{bmatrix}^{\top}$, with the reference input $\mathbf{u}^{\text{ref}} = \boldsymbol{0}$.
The vectors $\mathbf{s}_0$ and $\mathbf{s}^{\text{ref}}$ remain fixed throughout all learning episodes.
The RL stage cost $\ell$ used to evaluate the RL performance defined in Paragraph \ref{section:prerequisites} is
\begin{equation} \label{eq:RL_stage_cost}
	\ell = \begin{cases}
		\left\|\mathbf{s}_k - \mathbf{s}^{\text{ref}}\right\|_Q^2 + \left\|\mathbf{u}_k\right\|_2 & \text{if }\left\|\mathbf{s}_k-\mathbf{s}^{\text{ref}}\right\|_2 < d_t\\
		\left\|\mathbf{s}_k-\mathbf{s}^{\text{ref}}\right\|_Q^2 + \Upsilon(\mathbf{s}_k) & \text{if }\left\|\mathbf{s}_k-\mathbf{s}^{\text{ref}}\right\|_2 \geq d_t
	\end{cases},
\end{equation}
where $Q = \operatorname{diag}(1, 1, 0.1)$ is the objective weighting  matrix, $d_t = \SI{1.5}{\meter}$ is a constant distance threshold, and $\Upsilon(\mathbf{s}_k) = \sum_{n = 1}^{4} c\max\{0, \Xi(\mathbf{s}_k) + d_o\}$ is the obstacle penalty collision where $\Xi$ is as defined as in \eqref{eq:obstacle_constraint}, and $d_o = \SI{.35}{\meter}$ is a desired safe distance between the robot and the obstacle, and $c = 50$ is a penalty weight.
If a collision occurs, $\Upsilon(\mathbf{s}_k)$ adds a positive penalty to the RL stage cost $\ell$.

\begin{table}[!tb]
	\caption{Controller and algorithm configuration.}
	\label{algo_params}
	\centering
	\begin{tabular}{ll}
	\hline
		Parameter & Value\\
	\hline
		$\alpha$, $\beta$, $\zeta$  				      & $\num{e-7}$, $\num{e-8}$, $\num{e-2}$\\
		$\mathbf{w}_{\text{init}}$                  & $\num{e-4}$\\
		Robot $\boldsymbol{\theta}_{\text{init}}$             & $[1\ 1\ 0.05\ 0.05\ 0.05\ 1\ 1\ 0.1\ 0.001]$ \\
		Law of $\boldsymbol{\rho}$, $\gamma$                 & $\mathcal{N}(0, \num{1})$, $0.97$\\
		$N$, $K$, $N_{\text{ep}}$   		                 & $10$, $129$, $300$\\
		$\boldsymbol{\omega}$, $\boldsymbol{\omega}_f$			& $[100,100,100,100]^{\top}$\\

		Mini-batch size $\operatorname{card}(\mathcal{B})$               & $128$\\
		Hidden layers, neurons per layer                                & $2$, $64$\\
	\hline
	\end{tabular}
\end{table}

	\subsection{Simulation results}

To showcase the advantages of our algorithm, we train $\boldsymbol{\theta}$ for 300 episodes in four ways: 
one using the method from \cite{moradimaryamnegari2022model}, one using the method from \cite{moradimaryamnegari2023data}, and the last two using the proposed methods from paragraphs \ref{subsection:regularized_deep_expected_sarsa} and  \ref{subsection:GTD}. 
The numerical computation is performed using the Ipopt solver provided by the CasADi software framework \cite{Andersson2019} on a PC equipped of 16 GB of RAM.     

\begin{figure}[!tb]
	\centering
	\includegraphics[width = .9\columnwidth]{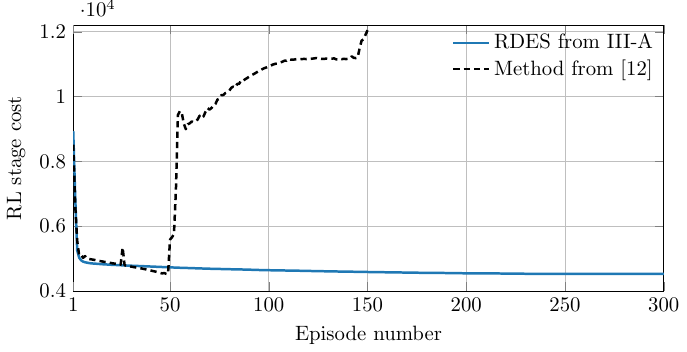}
	\caption{RL stage cost $\ell$ sum over learning episodes.}
	\label{fig:performance1}
\end{figure}

\begin{figure}[!tb]
	\centering
	\includegraphics[width = .9\columnwidth]{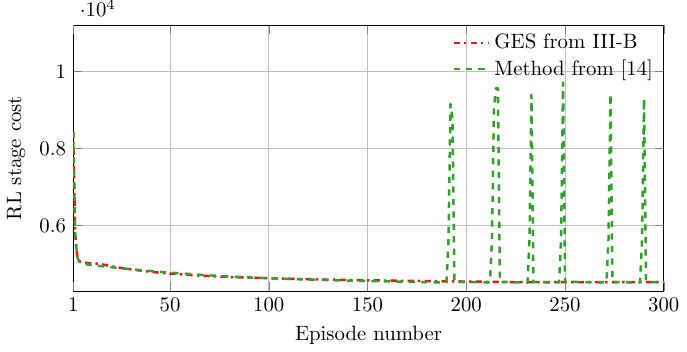}
	\caption{RL stage cost $\ell$ sum over learning episodes.}
	\label{fig:performance2}
\end{figure}

For fair comparison, we compare the deep ES method from \cite{moradimaryamnegari2023data} with our RDES method from paragraph \ref{subsection:regularized_deep_expected_sarsa}.
Indeed, for both these methods, only one NMPC must be solved to compute the current action-value function, whereas the subsequent action-value function is approximated by a neural network.
Fig. \ref{fig:performance1} shows the sum of RL stage cost across training episodes for both approaches. 
In the blue graph, we can observe that the RL performance decreases faster, with the learning process remaining stable throughout training episodes, highlighting the effect of adding the parametrization vector into the NN's training input for better approximation of the subsequent action-value function.
In contrast, the black dashed graph from \cite{moradimaryamnegari2023data} shows divergent RL performance, likely due to inaccuracies in approximating the subsequent action-value function.
 
To demonstrate the effectiveness of our second contribution from Paragraph \ref{subsection:GTD}, we compare our proposed GES with the ES method from \cite{moradimaryamnegari2022model}.
In both approaches, two NMPCs are solved to compute the TD error in \eqref{eq:sarsa_tderror}. 
As we already stated, conventional temporal-difference learning methods can become unstable or even diverge when nonlinear function approximation is involved. 
In Fig. \ref{fig:performance2}, we see that the blue graph reflects an unstable learning process as training episodes progress, attributed to the nonlinearities in the NMPC function approximator.
In contrast, the red graph displays stable RL performance, with the GES method converging toward a stable equilibrium point. 
This demonstrates the effectiveness of GTD learning, which is a true gradient descent method, ensuring convergence.

\begin{figure}[!tb]
	\centering
	\includegraphics[width = .8\columnwidth]{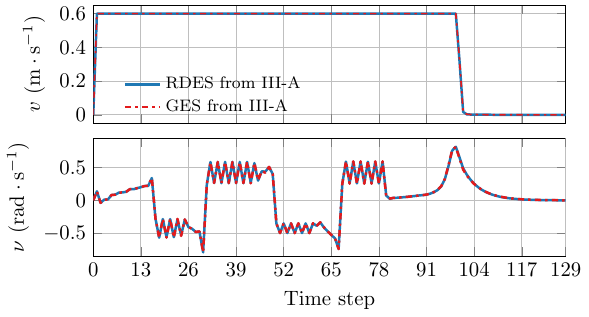}
	\caption{Robot control signals under the learned policies.}
	\label{fig:rob_control_signals}
\end{figure}

\begin{figure}[!tb]
	\centering
	\includegraphics[width = .9\columnwidth]{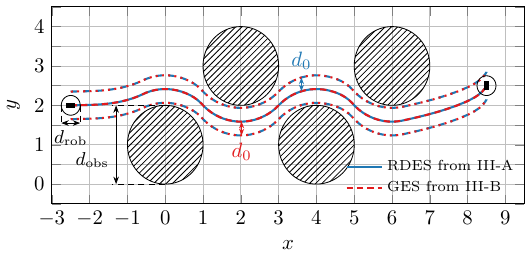}
	\caption{Robot states' trajectories under the learned policies.}
	\label{fig:robot_traj}
\end{figure}

\begin{table}[!tb]
	\caption{Training time comparison.}
	\label{table: Training time comparison}
	\centering
	\begin{tabular}{ll}
		\hline
			Method & Training time ($\si{\hour}$)\\
		\hline
			NMPC-based ES from \cite{moradimaryamnegari2022model} & $\approx$ 18\\
			NMPC-based GES from \ref{subsection:GTD} & $\approx$ 18\\
			NMPC-based RDES from \ref{subsection:regularized_deep_expected_sarsa} & $\approx$ 9\\
		\hline
	\end{tabular}
\end{table}

Table \ref{table: Training time comparison} presents the real-time training duration of the three methods for 300 learning episodes.
The deep ES method from \cite{moradimaryamnegari2023data} is excluded, as it diverges before reaching the 300th episode.
As shown, our RDES method completes the training episodes in half the time required for both the method from \cite{moradimaryamnegari2022model} and our GES method.

\section{Conclusion}

This paper proposes two reinforcement learning-based methods for tuning a nonlinear model predictive controller (NMPC).
In the first method, a parametrized NMPC is used as the current action-value function in a regularized deep Expected Sarsa (RDES) algorithm, while a neural network (NN) with a novel feature representation serves as the subsequent action-value function.
By incorporating the NMPC parameter vector into the NN's input training data, the learning process becomes more stable, and training time is reduced without compromising closed-loop performance.
The second method introduces a combination of NMPC-based gradient Expected Sarsa (GES) with convergence guarantees.
This approach stabilizes learning and ensures convergence to an equilibrium point, even in the presence of nonlinearities.
Finally, we find that both methods achieve the same optimal policy.
However, RDES reaches this outcome in half the training time, whereas GES provides convergence guarantees to a locally stable equilibrium.

In NMPC-based RL methods, adapting to dynamic environments, such as shifting obstacle positions during learning, depends on whether the changes are deterministic or stochastic.
Deterministic changes allow pre-scheduled updates to the nonlinear programming (NLP) problem, adjusting constraints and costs to account for updated obstacle positions, while the RL method adapts its value function and policy accordingly. 
For stochastic changes, performance metrics or real-time detection of obstacle movements trigger dynamic updates to the NLP scheme, modifying the NLP structure to match the current environment.
Meanwhile, for the RL component, one can consider adapting its learning rate to quickly update the value function and prioritizing recent data in the replay buffer to focus on the current environment's state.
This ensures the policy remains effective and the system handles environmental shifts robustly, whether changes are predictable or random.

\bibliographystyle{IEEEtran}
\bibliography{learning_based_nmpc}

\end{document}